\documentclass[11pt,twocolumn,showkeys]{revtex4-2}
\usepackage{relsize}
\usepackage{hyperref}
\usepackage{amsmath,amsfonts,amssymb}
\hypersetup{dvips,dvipdfm,colorlinks=true,urlcolor=magenta,filecolor=magenta,linktoc=page,citecolor=red,linkcolor=blue,bookmarks=true}
\usepackage{graphicx,epstopdf}
\begin{document}
	
	\title {Cyclic evolution of a two fluid diffusive Universe}
	\author{Subhayan Maity\footnote {maitysubhayan@gmail.com}}
	\affiliation{Department of Mathematics, Jadavpur University, Kolkata-700032, West Bengal, India.}

\begin{abstract}
 Complete scenario of cosmic evolution from emergent phase to late time acceleration (i.e. non-singular ever expanding Universe) is a popular preference in the recent cosmology. Yet one can't exclude the idea that other type of evolution pattern of the Universe may also be possible. Especially, the bouncing scenario  is  becoming a matter of interest now a days. The present work is an exhibition of such a different pattern of cosmic evolution where the evolution of Universe has been shown as a cyclic thermodynamic process. Under diffusion mechanism (non-equilibrium thermodynamic process), the cosmic evolution has been modelled as [ emergent $\rightarrowtail$ accelerated expansion $\rightarrowtail$ decelerated expansion $\rightarrowtail$ decelerated contraction $\rightarrowtail$ accelerated contraction $\rightarrowtail$ emergent] .

\end{abstract}

\maketitle





 \ 'Bulk flow' or 'advection' is the motion or flow of an entire substance from or to the system due to the pressure gradient (for example water flow out of tap). On the other hand, 'diffusion' is the gradual transport or dispersion of concentration within a body, due to the concentration gradient with no net movement of the substance from or to the system. The diffusion depends on the particle random walk motion and results in mixing or mass transport without any net bulk motion. Inhomogeneity of concentration within the system (it may be due to various process like bulk flow, particle creation-annihilation etc.) leads to the disturbance in the equilibrium condition among the different parts of the system. To restore the  equilibrium state, a pressure termed as 'diffusive pressure' is generated which drives the flow of molecules from higher to lower concentration region. Both advection and diffusion processes  are the transport phenomena and hence thermodynamically irreversible process \cite{MacDougall}. The combination of these two irreversible process is called 'convection'. Convection process results a net transport of mass and it occurs in a non-equilibrium system. So, to apply the laws of thermodynamics, one may assume the quasi-steady state where the process  changes with time very slowly and then the thermodynamic laws are applicable instantaneously \cite{MacDougall}.
\par Here, Universe is assumed as a system of two fluids. One is a diffusive fluid with a dissipative pressure called diffusion pressure ($\pi$). Another is non-diffusive fluid. In  cosmic diffusion process, the diffusion pressure deficiency leads to the  transport phenomena as wel as   it causes a continuous expansion of the Universe so that it can never be in equilibrium . Hence, the cosmic diffusion process can be described as diffusion within a continuously evolving isolated system. So, the Fokker-Planck equation (which describes the continuity condition of diffusion process)  must be modified for a continuously expanding system.    \par At the present context,  the energy momentum tensor $\left (T^{(d)}_{\mu \nu} \right )$ of the diffusive fluid is  chosen in the form,

\begin{figure}[h]
	\begin{minipage}{0.5\textwidth}
		\centering
		\includegraphics*[width=0.6\linewidth]{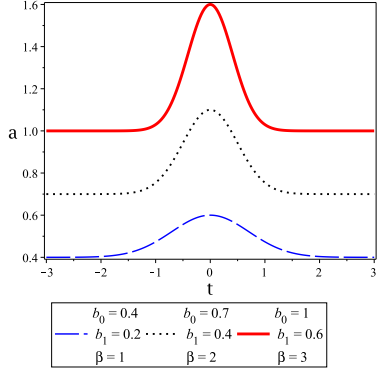}\\
		(a)
	\end{minipage}
	\begin{minipage}{0.5\textwidth}
		\centering
		\includegraphics*[width=0.6\linewidth]{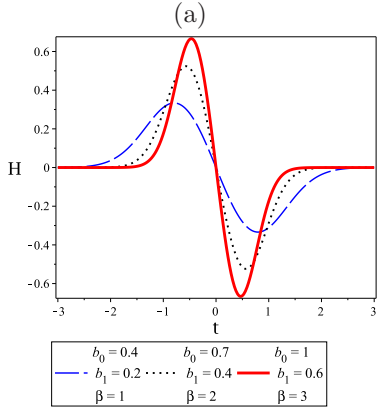}\\
		(b)
		
	\end{minipage}

	\begin{center}
		\caption{Variation of different cosmological  parameters namely ~~ (a)Scale factor ($a$) (b) Hubble parameter $H$ for the set of values $(b_0,b_1,\beta)=(0.4,0.2,1)=(0.7,0.4,2)=(1,0.6,3)$ respectively.}
	\end{center}
	\label{fig1}
\end{figure}

\begin{equation}
	T^{(d)}_{\mu \nu} = (P+\rho ^{(0)})u_{\mu}u_{\nu}- P g_{\mu \nu} ,     \label{1}
\end{equation}  
where $\rho ^{(0)}$ is the density of the diffusive fluid and the total pressure, $P=P^{(0)}+\pi$, $P^{(0)}$ being the thermodynamic pressure of the  fluid. 

The other fluid has the energy momentum tensor as, 
\begin{equation}
	\tilde{T_{\mu \nu}} =(\tilde{P}+\tilde{\rho})u_{\mu}u_{\nu}- \tilde{P} g_{\mu \nu}   \label{2}
\end{equation}

Due to the mutual interaction of the two fluids, the energy momentum tensor of individual fluid will not be conserved and the dissipation of the two fluids can be expressed as,

\begin{subequations}
	\begin{align}
		\nabla _{\mu} T_{(d)}^{\mu \nu} &= Q_{(d)}^{\nu} ~~~ \mbox{and}    \label{3a} \\
		\nabla _{\mu} \tilde{T}^{\mu \nu}& = \tilde{Q}^{\nu},   \label{3b}
	\end{align}	
\end{subequations}

\begin{figure}[h]
	\begin{minipage}{0.5\textwidth}
		\centering
		\includegraphics*[width=0.6\linewidth]{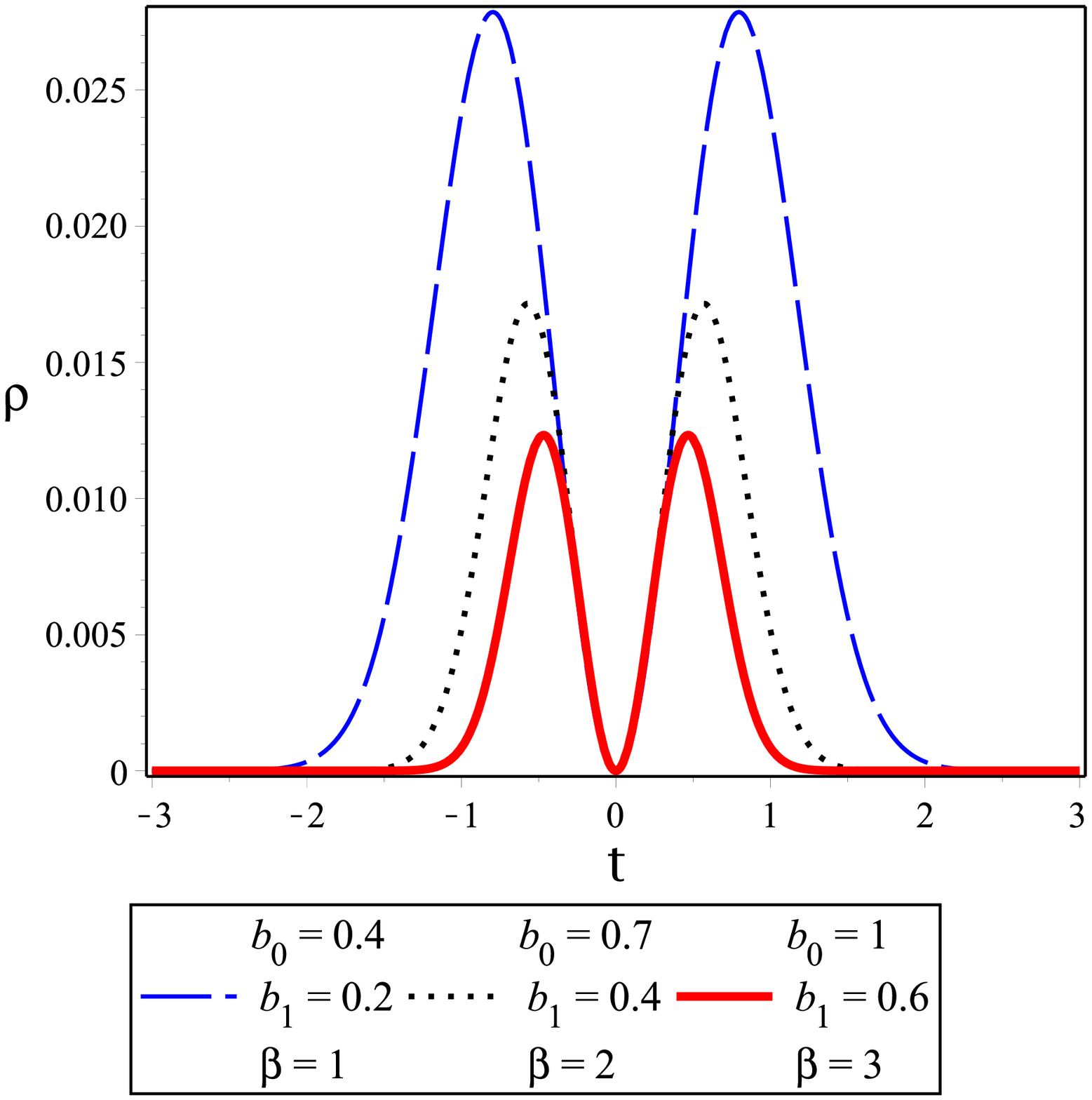}\\
		(a)
	\end{minipage}
	\begin{minipage}{0.5\textwidth}
		\centering
		\includegraphics*[width=0.6\linewidth]{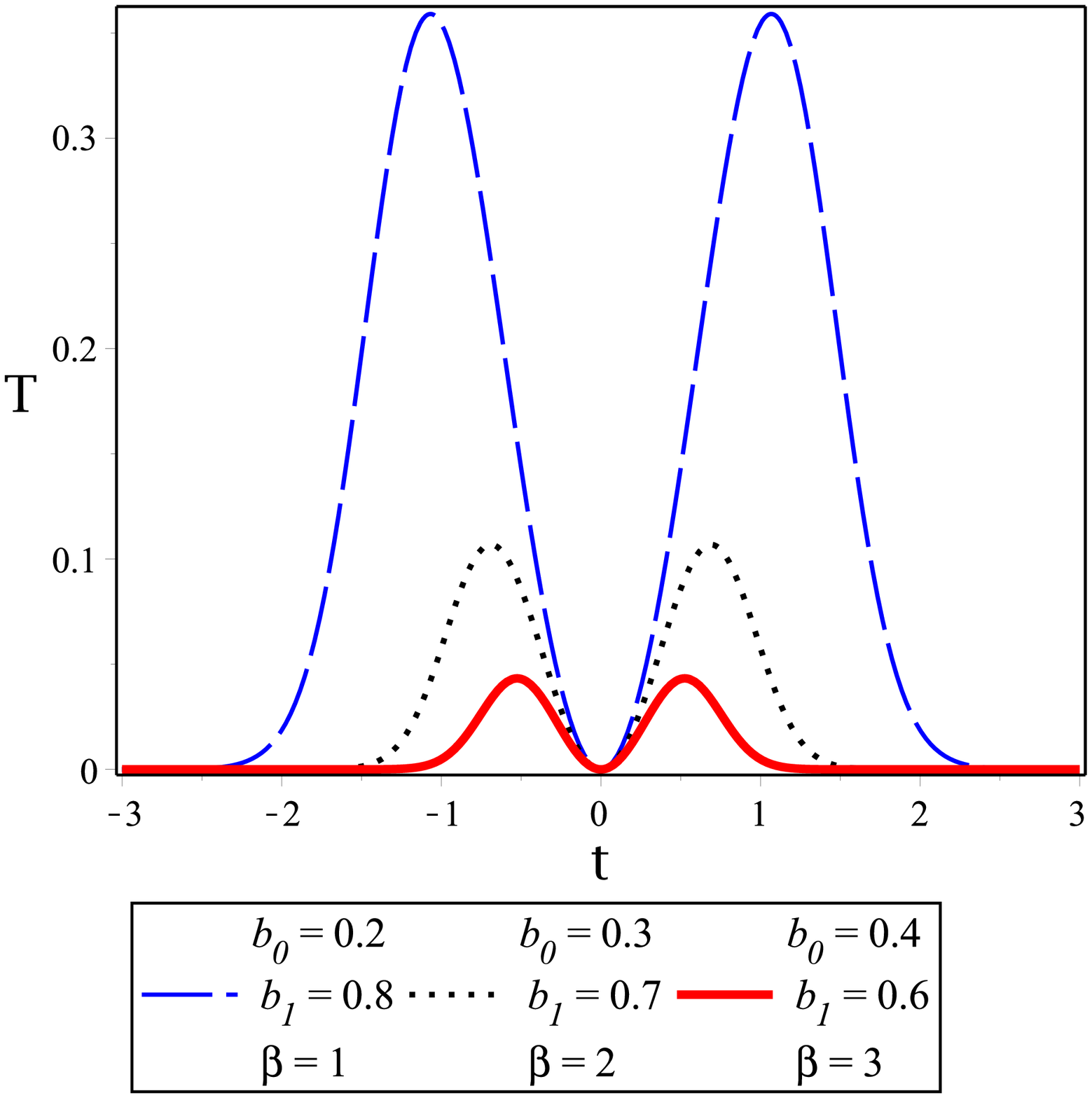}\\
		(b)
		
	\end{minipage}

	\begin{center}
		\caption{Variation of different thermodynamic  parameters namely ~~ (a)Energy density ($\rho$) (b) Temperature $T$ for the set of values $(b_0,b_1,\beta)=(0.4,0.2,1),(0.7,0.4,2),(1,0.6,3)$ respectively.}
	\end{center}
	\label{fig2}
\end{figure}

where $ Q_{(d)}^{\nu}$ and $\tilde{Q}^{\nu}$ are the interaction terms for the two fluids respectively.  In an isolated universe, the total energy-momentum tensor will be conserved and hence one finds, $ Q_{(d)}^{\nu}=-\tilde{Q}^{\nu}$. The interaction between two fluids can be considered as the particle creation-annihilation mechanism\cite{Chakraborty:2014oya}. \par Asper Bianchi identity, it is obvious that the effective density $\rho =\rho^{(0)}+\tilde{\rho}$ and the effective pressure $P_{\mbox{eff}}=P^{(0)}+\tilde{P}$
\begin{equation}
	\nabla _{\mu} T_{(d)}^{\mu \nu}+\nabla _{\mu} \tilde{T}^{\mu \nu}=0   \label{4}
\end{equation} 
In the frame work of general theory of relativity, the equation (\ref{4}) can be simplified for a homogeneous, isotropic and flat FLRW universe as,
\begin{equation}
	\frac{\partial \rho^{(0)}}{\partial t} +3 H(P^{(0)}+\pi+\rho^{(0)}) +\frac{\partial \tilde{\rho}}{\partial t} + 3 H (\tilde{P}+\tilde{\rho})=0,   \label{5}
\end{equation}
where $H=\frac{\dot{a}}{a}$ is the Hubble parameter of the universe. \par Now this combination of two fluids can be treated as a dissipating single fluid system with. The continuity equation for this system can be expressed as, 

\par Considering the two fluids as the barotropic fluids with barotropic index $\omega _0$ and $\tilde{\omega}$ respectively (i.e. $P^{(0)}=\omega _0 \rho ^{(0)}$ and $\tilde{P}=\tilde{\omega} \tilde{\rho}$), one can effectively write, $P_{\mbox{eff}}=\omega _0 \rho ^{(0)}+\tilde{\omega} \tilde{\rho}= \omega \rho$, where $\omega$ is the equivalent barotropic index of the combination of two fluids. 
Again for a diffusive system, one can estimate the conservation equation as \cite{Haba:2016bpt,Maity:2019knj,new},
\begin{equation}
	\frac{\partial \rho}{\partial t}+3 H(1+\omega)\rho =\gamma a^{-3} ,    \label{7}
\end{equation}
with $\gamma$ as diffusion parameter. The above equation (\ref{7}) is the modified form of Fokker - Planck equation in such model. Comparing equation (\ref{6}) with equation (\ref{7}), one finds the expression for diffusive pressure $\pi$ in terms of diffusion parameter $\gamma$ as, 
\begin{equation}
	\pi =-\gamma \frac{a^{-3}}{3H}.   \label{8}
\end{equation}
The corresponding Friedmann equations can be found in the forms,
\begin{subequations}
	\begin{align}
		3 H^2&=K \rho    \label{9a}  \\
		-2 \dot{H}&=K(P_{\mbox{eff}}+\pi+\rho) ,   \label{9b}
	\end{align}
\end{subequations}  with $K$ as a constant. Eliminating $K$ from Friedmann equations, one obtains the evolution equation of the universe as, 
\begin{equation}
	2 \dot{H}+3(1+\omega)H^2 =\frac{\gamma a^{-3}}{\rho}H .  \label{10}
\end{equation}
Again the solution of the conservation equation(\ref{7}) yields,
\begin{equation}
	\rho =a^{-3(1+\omega)}\left(\rho _0 +\int_{t_0}^{t} \gamma  a^{3\omega} dt   \right),   \label{11}
\end{equation}
where $\rho_0$ is the density at reference epoch of time $t_0$ with  $a(t_0)=1$ .
Substituting the value of $\rho$ from equation (\ref{11}), in the equation (\ref{10}), one obtains the solution for energy density in-terms of Hubble parameter as,
\begin{equation}
	\rho =\rho _0 \left (\frac{H}{H_0} \right )^2   \label{12}
\end{equation}
with $H_0$ is the value of Hubble parameter at $t=t_0$.

Under such diffusion process, the universe undergoes entropy production (as diffusion is a thermodynamically irreversible process) and the cosmic fluids suffer from thermal instability in arbitrary condition. Hence  one has to consider the instantaneous thermal equilibrium state of the universe with an instantaneous equilibrium temperature $(T)$. According to non-equilibrium thermodynamics, the evolution of instantaneous equilibrium temperature $T$ can be obtained through the relation \cite{Chakraborty:2014oya,Bhandari:2018fon},
\begin{equation}
	\frac{\dot{T}}{T}+\omega \left (3H+\frac{\pi}{1+\omega}\right ) =0  . \label{13}
\end{equation}
\begin{figure}[h]
	\begin{minipage}{0.5\textwidth}
		\centering
		\includegraphics*[width=0.6\linewidth]{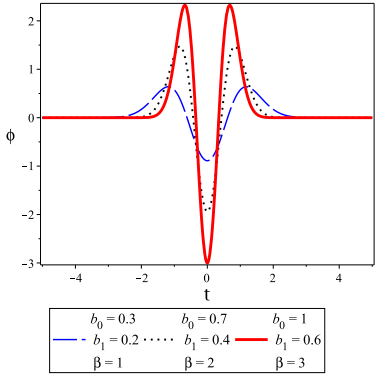}
		
	\end{minipage}

	\begin{center}
		\caption{Variation of  $\phi=h^{-1}$ with time $t$.}
	\end{center}
	\label{fig3}
\end{figure}

Here in this context, one can write the evolution of temperature as, 
\begin{equation}
	\frac{\dot{T}}{T}+\omega \left (3H-\frac{\gamma}{(1+\omega)\rho a^3}\right ) =0   \label{14}
\end{equation}

The dissipative term $\frac{\gamma a^{-3}}{\rho} H$ of the evolution equation (\ref{10}) of the Universe depends on the thermodynamic behaviour of the diffusive fluid. The enthalpy of the diffusive $h$ is related to the energy density $\rho$ as, 
\begin{equation}
	h=E+PV=(1+\omega)\rho V_0 a^3 .   \label{15}
\end{equation}
Hence, equation(\ref{10}) can be found in terms of enthalpy as,
\begin{equation}
	2 \dot{H}+3(1+\omega)H^2 =3\eta (1+\omega)H h^{-1} , \label{16}
\end{equation}
with $\eta =\frac{1}{3}\gamma V_0$, another constant.

Second law of thermodynamics insists that any system interacting with thermal energy converts a fraction of heat into the corresponding amount of work i.e. any thermodynamic  system has the heat engine property with efficiency $\eta < 1$. Carnot's first theorem states that reversible heat engine possesses the maximum possible efficiency and generally it follows a cyclic thermodynamic process between two heat reservoirs ($T_1$ and $T_2$, $T_1>T_2$) (say).The Carnot cycle is the most well known reversible engine with efficiency $\eta =1-\frac{T_2}{T_1}$.
Again asper Carnot's theorem, all reversible engines are equally efficient. 
Hence if one has the motivation to predict the cosmic evolution pattern from thermodynamic point of view then  it may be a quite wise decision to modell the Universe as a  reversible heat engine (not essentially the Carnot engine) . 

The cosmic heat engine clearly follows the cyclic and  bouncing evolution pattern and it can be assumed as (Emergent stage $\rightarrowtail$ Inflation $\rightarrowtail$ Decelerating expansion $\rightarrowtail$ Decelerating contraction $\rightarrowtail$ Inflationary contraction $\rightarrowtail$ Emergent) . Now in a bouncing Universe,  near emergent stage\cite{Banerjee:2007qi,Bhattacharya:2016env,Bose:2020xml,Chakraborty:2014ora,Ellis:2002we,Ellis:2003qz,Guendelman:2014bva}, $t\rightarrow \pm \infty ,a \rightarrow b_0 ( \mbox{a constant}), H\rightarrow 0 , h \sim $ constant. Also at, $t=\tilde{t}, H = 0, \gamma = 0$, where $\tilde{t}$ is the transition time epoch from expansion to contracting phase.So, the primitive choice (after trials with several different choices ) is taken as,
\begin{equation}
	h^{-1}  = \alpha _1  H^2 + \alpha _2  \frac{H}{t-\tilde{t}} + \alpha _3  \frac{b_0}{a} (t-\tilde{t} ~)H    \label{17}
\end{equation} where $\alpha _1, \alpha _2$ and $\alpha _3$ are constants. For simplicity of the calculation, one takes 
\par  $\alpha _1 =3$ and $\alpha _2 =\frac{-2}{1+\omega} ,\alpha _3=-\frac{4 \beta}{1+\omega}$ with $\beta $, a constant.

Then, the evolution equation (\ref{16}) takes  the form
\begin{equation}
	2 \dot{H}+4 \beta \left[(1-\frac{b_0}{a})+\frac{b_0}{a}(t-\tilde{t})H\right ]=0 . \label{18}
\end{equation}

The equation (\ref{18}) yields the solution,

in the simplified forms,
\begin{eqnarray}
	H= -2 \beta (t-\tilde{t}~) \left (1- \frac{b_0}{a} \right )    \label{19} \\
	a= b_0 + b_1 \exp \left [ -\beta (t-\tilde{t}~)^2     \right ]   \label{20}
\end{eqnarray}   where $b_1$ is also a constant.  Hence the deceleration parameter is found as,
\begin{equation}
	q=\frac{a}{a-b_0}\left [\frac{1}{2 \beta}.\frac{1}{(t-\tilde{t})^2}-1 \right ]         \label{21}
\end{equation} 
which suggests that, with in the epoch range :  $\tilde{t}-\frac{1}{\sqrt{2\beta }} \leq t \leq \tilde{t}+\frac{1}{\sqrt{2\beta }} $, the deceleration occurs and beyond this range, acceleration occurs. The other two thermodynamic parameters are as,
\begin{eqnarray}
	\rho \sim \left[(t-\tilde{t}~) \left(1-\frac{b_0}{a}\right ) \right ]^2   \label{22}                    \\
	T \sim \left[(t-\tilde{t}~) \left(1-\frac{b_0}{a} \right ) \right ]^{\frac{2 \omega}{1+\omega}}.    \label{23}
\end{eqnarray}

The variation of different parameters has been presented graphically in FIG. $1,2$ and $3$. In this cyclic process, Universe starts from emergent phase and it follows successive accelerated and decelerated expansions. Then successively decelerated contraction and accelerated contraction occur and  finally it leads to the emergent phase again. Intersetingly, the initial and final emergent phases are thermodynamically identical but differnt in cosmic perspective (initial emergent phase has positive Hubble parameter but the final one possesses negative Hubble parameter). So this process can  be described as a cyclic process thermodynamically only. Conclusively, it can be speculated if a worm hole exists between the initial and final emergent phases then this cycle will be repeated for ever.

\section*{Acknowledgements}  The author SM acknowledges UGC for
awarding Research fellowship and also thanks Prof.Subenoy Chakraborty, Dept. of Mathematics, J.U. for his valuable suggestions.

\end{document}